\documentclass[aps,preprint]{revtex4}%
\usepackage{amsfonts}
\usepackage{amsmath}
\usepackage{amssymb}
\usepackage{graphicx}%
\providecommand{\U}[1]{\protect\rule{.1in}{.1in}}
\begin{document}
\title{Engineering phonon-photon interactions with a driven trapped ion in a cavity }
\author{R. L. Rodrigues, M. H. Y. Moussa, and C. J. Villas-B\^{o}as}
\affiliation{Departamento de F\'{\i}sica, Universidade Federal de S\~{a}o Carlos, P.O. Box
676, S\~{a}o Carlos\textit{, 13565-905, S\~{a}o Paulo, }Brazil}

\begin{abstract}
We show how to generate quadratic and bi-quadratic phonon-photon interactions
through a driven three-level ion inside a cavity. With such a system it is
possible to squeeze the cavity-field state, the ion motional state or even the
entangled phonon-photon state. We present a detailed analysis of the
cavity-field squeezing process, distinguishing three different regimes of this
amplification mechanism: the subcritical, critical, and supercritical regimes,
which depend, apart from the coupling parameters, on the excitation of the
vibrational state. As an application of the engineered Hamiltonians, we show
how to implement a Fock-state filter for the vibrational mode. New aspects of
the technique of adiabatic elimination emerge in this analysis.

\end{abstract}

\maketitle

\section{Introduction}

Together with cavity quantum electrodynamics (QED) and manipulation of light
states through linear and nonlinear optical elements, the physics of trapped
ions is a major ingredient of the quantum information theory research scene.
The experimental achievements aligned with the theoretical propositions in
these domains of quantum optics have contributed significantly to the
insertion of quantum information theory in virtually all the areas of nowadays
physics. The possibility to insert a trapped ion inside a cavity to manipulate
phonon-photon interaction has been raised from the very beginning of the
period of experimental accomplishments in cavity QED and trapped ions
\cite{Walls}. Since then, the problem of phonon-photon interaction with a
trapped ion inside a cavity has attracted attention, owing to its application
to quantum logic operation \cite{Zou,Semiao,Feng}, to the translation of
phonon to photon statistics or the transference of squeezing from phonons to
photons \cite{Orszag}, and to the study of the dynamics of the interaction
between a cavity field and the motional degrees of freedom of a trapped ion
\cite{Vogel}. A scheme for quantum swapping between vibrational and cavity
field states has also been presented \cite{Zagury}, not to mention that the
engineering of quantum states in such a system, specially of entangled atomic
motion and cavity field \cite{DiFidio}, is considered in all these references.
Parallel to the study and applications of phonon-photon interaction, the major
mechanisms of decoherence in ionic traps have been experimentally analyzed
\cite{Wineland} and modeled \cite{Vogel1,Serra,Budini}. The knowledge acquired
over the last few years about dissipative mechanisms in both systems, high-Q
cavities and ionic traps, has resulted in protocols for quantum-state
protection in ionic traps \cite{Luis} and cavity QED \cite{Celso}.

In the present paper we are interested in engineering phonon-photon
interactions with a trapped ion inside a cavity. The program of engineering
Hamiltonians has become a major concern in quantum information research:
beyond the need for quantum state preparation, a given logical operation
requires specific interactions between the subsystems comprising the quantum
bits. Recent work has been devoted to engineering bilinear interactions in
two-mode cavity QED; specifically, parametric up- and down-conversion
operations were accomplished through the dispersive interactions of the cavity
modes with a single three-level-driven atom which works as a nonlinear medium
\cite{Rapid,Roberto,Celso1,Guzman}. Here, a three-level trapped ion
interacting simultaneously with a classical field and a single cavity mode
will be treated by the adiabatic approximation technique. New aspects of this
approximation are revealed through our analysis which handles both the weak
and the strong-amplification regimes of the classical field.

\section{The Model}

The energy diagram of the three-level trapped ion, sketched in Fig. 1, is in
the ladder configuration, where the ground $\left\vert g\right\rangle $ and
excited $\left\vert e\right\rangle $ states, with transition frequency
$\omega_{0}$, are coupled through an intermediate level $\left\vert
i\right\rangle $. The cavity mode of frequency $\omega$ is tuned to the
vicinity of both dipole-allowed transitions, $\left\vert g\right\rangle $
$\leftrightarrow$ $\left\vert i\right\rangle $ and $\left\vert e\right\rangle
$ $\leftrightarrow$ $\left\vert i\right\rangle $, with coupling constants
$\lambda_{1}$ and $\lambda_{2}$ and detunings $\delta_{1}=$ $\Delta+\delta$
and $\delta_{2}=\Delta-\delta$, respectively, where $\delta=\omega
_{0}/2-\omega$. Simultaneously to the cavity mode, a classical field is
assumed to drive resonantly the dipole-forbidden atomic transition $\left\vert
g\right\rangle $ $\leftrightarrow$ $\left\vert e\right\rangle $ with coupling
constant $\Omega$ \cite{Forbidden}.{\Large \ }The Hamiltonian which describes
this system is given by $H=H_{0}+V(t)$, where (with $\hbar=1$):%

\begin{subequations}
\begin{align}
H_{0}  &  =\omega a^{\dagger}a+\nu b^{\dagger}b+\left(  \omega+\delta\right)
\left(  \sigma_{ee}-\sigma_{gg}\right)  +\Delta\sigma_{ii},\label{1a}\\
V(t)  &  =\left(  \lambda_{1}\sigma_{gi}+\lambda_{2}\sigma_{ie}\right)
\left(  a^{\dagger}+a\right)  \sin\left[  \eta\left(  b^{\dagger}+b\right)
+\varphi\right] \nonumber\\
&  +\Omega\exp\left[  -2i\left(  \omega+\delta\right)  t\right]  \sigma
_{eg}\exp\left[  i\eta_{L}\left(  b^{\dagger}+b\right)  \right]
+\mathrm{h.{c.}} \label{1b}%
\end{align}
with $a^{\dagger}$ ($a$) and $b^{\dagger}$ ($b$) standing for the creation
(annihilation) operators of the quantized cavity mode and the one-dimensional
trapped motion of frequency $\nu$, while $\sigma_{rs}\equiv\left\vert
r\right\rangle \left\langle s\right\vert $ ($r,s$ being the atomic states
$g,e,i$). The Lamb-Dicke parameter $\eta_{L}=$ $k_{L}/\sqrt{2m\nu}$ $\left(
\eta=k/\sqrt{2m\nu}\right)  $ follows from the interaction of the ion with the
classical (quantum) radiation field characterized by the wave vector
$k_{L}=\omega_{L}/c$ $\left(  k=\omega/c\right)  $. Finally, $\varphi$
accounts for the relative position of the ion in the cavity standing wave,
such that $\varphi=0$ ($\pi/2$) corresponds to an ion centered at a node
(anti-node) of the standing wave. The phase accounting for the relative
position of the ion with respect to the classical field is incorporated into
the complex constant $\Omega=\left\vert \Omega\right\vert \operatorname*{e}%
\nolimits^{-i\phi}$. After the unitary transformations $U_{0}=\exp\left(
-iH_{0}t\right)  $ and $U_{1}=\exp\left(  i\Delta\sigma_{ii}t\right)  $,
associated with the interaction picture and a frame rotating with frequency
$\Delta$, respectively, the Hamiltonian becomes $\mathcal{H}=U_{1}^{\dagger}$
$U_{0}^{\dagger}HU_{0}$ $U_{1}-H_{0}+\Delta\sigma_{ii}$. Assuming from here on
that $\omega\gg\nu,\delta,\Delta,\lambda_{1},\lambda_{2}$ and keeping terms of
order of $\eta^{2}$ within the Lamb-Dicke limit $\eta^{3}\ll1$, we obtain
within the rotating-wave approximation (neglecting terms rotating with
frequency of the order of $2\omega$):%

\end{subequations}
\begin{equation}
\mathcal{H}(t)=\left[  \left(  \lambda_{1}\sigma_{gi}+\lambda_{2}\sigma
_{ie}\right)  a^{\dagger}\Lambda\left(  b,b^{\dagger};t\right)  +\Omega
\sigma_{eg}\Sigma\left(  b,b^{\dagger};t\right)  +\mathrm{h.{c.}}\right]
+\Delta\sigma_{ii}\mathrm{{,}} \label{2}%
\end{equation}
where the time-dependent functions for the trapped-motion operators are given by%

\begin{subequations}
\begin{align}
\Lambda\left(  b,b^{\dagger};t\right)   &  =\exp\left(  -i\delta t\right)
\exp\left(  -i\nu b^{\dagger}bt\right)  \sin\left[  \eta\left(  b^{\dagger
}+b\right)  +\varphi_{c}\right]  \exp\left(  i\nu b^{\dagger}bt\right)
\mathrm{{,}}\label{3a}\\
\Sigma\left(  b,b^{\dagger};t\right)   &  =\exp\left(  -i\nu b^{\dagger
}bt\right)  \exp\left[  i\eta_{L}\left(  b^{\dagger}+b\right)  \right]
\exp\left(  i\nu b^{\dagger}bt\right)  \mathrm{{.}} \label{3b}%
\end{align}

Defining a new basis for the atomic states $\left\{  \left\vert i\right\rangle
,\left\vert \pm\right\rangle =\left[  \left\vert e\right\rangle \pm
\operatorname*{e}\nolimits^{i\phi}\left\vert g\right\rangle \right]  /\sqrt
{2}\right\}  $ \cite{Solano}, composed of eigenstates of the atomic
Hamiltonian\textbf{\ }$\left\vert \Omega\right\vert \left(  \operatorname*{e}%
\nolimits^{-i\phi}\sigma_{eg}+\operatorname*{e}\nolimits^{i\phi}\sigma
_{eg}\right)  $, and assuming the Lamb-Dicke-like limit $\eta_{L}\ll1$, such
that $\Sigma\left(  b,b^{\dagger};t\right)  \approx1$, we obtain%

\end{subequations}
\begin{align}
\mathcal{H}(t)  &  =\frac{1}{\sqrt{2}}\left[  \left(  \lambda_{1}%
\operatorname*{e}\nolimits^{-i\phi}a^{\dagger}\Lambda+\lambda_{2}^{\ast
}a\Lambda^{\dagger}\right)  \sigma_{+i}-\left(  \lambda_{1}\operatorname*{e}%
\nolimits^{-i\phi}a^{\dagger}\Lambda-\lambda_{2}^{\ast}a\Lambda^{\dagger
}\right)  \sigma_{-i}+\mathrm{h.{c}}\right] \nonumber\\
&  +\Delta\sigma_{ii}+\left\vert \Omega\right\vert \left(  \sigma_{++}%
-\sigma_{--}\right)  \mathrm{{.}} \label{4}%
\end{align}

\subsection{The Adiabatic Approximation}

Next, we proceed with a two-step approach for the adiabatic elimination of
both transitions $\left\vert +\right\rangle $ $\leftrightarrow$ $\left\vert
i\right\rangle $ and $\left\vert -\right\rangle $ $\leftrightarrow\left\vert
i\right\rangle $. In the first step we write, from the Liouville-von Neumann
equation $\overset{.}{\rho}=-i\left[  \mathcal{H}(t),\rho\right]  $, the
operator $\overset{.}{\rho}_{rs}$associated with the transition $\left\vert
r\right\rangle $ $\leftrightarrow$ $\left\vert s\right\rangle $ (with
$r,s=+,-,i$). Imposing the condition $\overset{.}{\rho}_{+i}=\overset{.}{\rho
}_{-i}=0$, we obtain the adiabatic solutions for both operators $\rho_{+i}$
and $\rho_{-i}$. The substitution of these solutions back into $\overset
{.}{\rho}_{++}$, $\overset{.}{\rho}_{--}$, and $\overset{.}{\rho}_{ii}$
results in the evolution operators for the probabilities of measuring the
electronic states $\left\vert +\right\rangle $, $\left\vert -\right\rangle $,
and $\left\vert i\right\rangle $, respectively. Next, in the second step,
assuming that the Hamiltonian under the adiabatic approximation (superscript
$\emph{A}$) is given by%

\begin{equation}
\mathcal{H}(t)=\mathcal{H}_{++}^{A}\sigma_{++}+\mathcal{H}_{--}^{A}\sigma
_{--}+\mathcal{H}_{ii}^{A}\sigma_{ii}+\left(  \mathcal{H}_{+-}^{A}\sigma
_{+-}+\mathrm{h.{c.}}\right)  \mathrm{{,}} \label{5}%
\end{equation}
in which the unwanted transitions are missing, we employ again the
Liouville-von Neumann equation to write new transition operators $\overset
{.}{\rho^{A}}_{rs}$. Comparing the operators $\overset{.}{\rho}_{++}^{A}$,
$\overset{.}{\rho}_{--}^{A}$, and $\overset{.}{\rho}_{ii}^{A}$, with those
obtained previously in the first step, $\overset{.}{\rho}_{++}$, $\overset
{.}{\rho}_{--}$, and $\overset{.}{\rho}_{ii}$, we finally obtain the
Hamiltonian terms $\mathcal{H}_{++}^{A}$, $\mathcal{H}_{--}^{A}$,
$\mathcal{H}_{ii}^{A}$, $\mathcal{H}_{+-}^{A}$, and $\mathcal{H}_{-+}^{A}$.
The validity of the adiabatic approximation follows from that of the adiabatic
solutions
\begin{equation}
\left\vert \Delta\pm\left\vert \Omega\right\vert \right\vert \gg\left\vert
\lambda_{1}\right\vert ,\left\vert \lambda_{2}\right\vert ,\delta\mathrm{{,}}
\label{6}%
\end{equation}
which leads to two different regimes of parameters: $i)$ the
weak-amplification regime, where $\Delta\gg\left\vert \Omega\right\vert
,\left\vert \lambda_{1}\right\vert ,\left\vert \lambda_{2}\right\vert ,\delta
$, and $ii)$ the strong-amplification regime, where $\left\vert \Omega
\right\vert \gg\Delta,\left\vert \lambda_{1}\right\vert ,\left\vert
\lambda_{2}\right\vert ,\delta$. We note that the technique of adiabatic
elimination, as used in the literature to date, applies only to the
weak-amplification regime. The strong-amplification regime defined above is
another, complementary aspect of adiabatic elimination. The resulting
Hamiltonian terms are given by
\begin{equation}
\mathcal{H}_{\ell k}^{A}=\left[  \omega_{\ell k}a^{\dagger}a+\chi_{\ell
k}+\left(  \xi_{\ell k}\operatorname*{e}\nolimits^{-i2\delta t}\left(
a^{\dagger}\right)  ^{2}+\mathrm{H.{c.}}\right)  \right]  VV^{\dagger
}\mathrm{{,}} \label{7}%
\end{equation}
with $\ell,k=+,-,i$, $V=U_{0}^{\dagger}\sin\left[  \eta\left(  b^{\dagger
}+b\right)  \right]  $ $U_{0}$; defining, in the weak ($w$) and strong
($s$)-amplification regimes, effective frequencies ($\omega_{w}$, $\omega_{s}%
$), coupling strengths ($\xi_{w}$, $\xi_{s}$), and energy shifts ($\chi_{w}$,
$\chi_{s}$), as%
\begin{align}
\omega_{w}  &  =\frac{\left\vert \lambda_{1}\right\vert ^{2}+\left\vert
\lambda_{2}\right\vert ^{2}}{\Delta}\text{ \ \ , \ \ }\omega_{s}%
=\frac{\left\vert \lambda_{1}\right\vert ^{2}+\left\vert \lambda
_{2}\right\vert ^{2}}{\left\vert \Omega\right\vert }\text{,}\nonumber\\
\xi_{w}  &  =\frac{\lambda_{1}\lambda_{2}\operatorname*{e}\nolimits^{-i\phi}%
}{\Delta}\text{ \ \ \ \ \ \ , \ \ }\xi_{s}=\frac{\lambda_{1}\lambda
_{2}\operatorname*{e}\nolimits^{-i\phi}}{\left\vert \Omega\right\vert
}\text{,}\label{8}\\
\chi_{w}  &  =\frac{\left\vert \lambda_{1}\right\vert ^{2}}{\Delta}\text{
\ \ \ \ \ \ \ \ \ \ \ \ , \ \ }\chi_{s}=\frac{\left\vert \lambda
_{1}\right\vert ^{2}}{\left\vert \Omega\right\vert }\text{,}\nonumber
\end{align}
the Hamiltonian parameters read, in the weak-amplification regime:%
\begin{align}
\omega_{ii}  &  \sim\omega_{w}\text{ \ \ \ \ \ \ \ \ \ \ \ \ \ , \ \ \ }%
\chi_{ii}\sim\chi_{w}\text{ \ \ \ \ \ \ \ \ \ \ \ , \ \ }\xi_{ii}\text{ }%
\sim-\frac{\left\vert \Omega\right\vert }{\Delta}\xi_{w}\text{ ,}\nonumber\\
\omega_{++}  &  \sim-\frac{1}{2}\omega_{w}\text{ \ \ \ \ \ \ \ \ , \ \ }%
\chi_{++}\sim-\frac{\left\vert \lambda_{2}\right\vert ^{2}}{2\left\vert
\lambda_{1}\right\vert ^{2}}\chi_{w}\text{ , \ \ }\xi_{++}\sim-\frac{1}{2}%
\xi_{w}\text{ \ ,}\nonumber\\
\omega_{--}  &  \sim\omega_{++}\text{ \ \ \ \ \ \ \ \ \ \ \ , \ \ }\chi
_{--}\sim\chi_{++}\text{ \ \ \ \ \ \ \ \ \ , \ \ }\xi_{--}\sim-\xi_{++}\text{
\ ,}\label{9}\\
\omega_{+-}  &  \sim\frac{\left\vert \lambda_{1}\right\vert ^{2}-\left\vert
\lambda_{2}\right\vert ^{2}}{2\Delta}\text{ , \ \ }\chi_{+-}\sim\chi
_{++}\text{ \ \ \ \ \ \ \ \ \ , \ \ }\xi_{+-}\sim\xi_{++}\text{\ \ \ \ \ ,}%
\nonumber
\end{align}

and in the strong-amplification regime:%
\begin{align}
\omega_{ii}  &  \sim-\frac{\Delta}{\left\vert \Omega\right\vert }\omega
_{s}\text{ \ , \ \ }\chi_{ii}\text{ }\sim-\frac{\Delta}{\left\vert
\Omega\right\vert }\chi_{s}\text{ \ \ \ \ , \ \ }\xi_{ii}\text{ }\sim-\xi
_{s}\text{ \ \ ,}\nonumber\\
\omega_{++}  &  \sim-\frac{1}{2}\omega_{s}\text{ \ \ \ , \ \ }\chi_{++}%
\sim\frac{\left\vert \lambda_{2}\right\vert ^{2}}{2\left\vert \lambda
_{1}\right\vert ^{2}}\chi_{s}\text{ \ , \ \ }\xi_{++}\sim\frac{1}{2}\xi
_{w}\text{ ,}\nonumber\\
\omega_{--}  &  \sim-\omega_{++}\text{ \ \ \ , \ \ }\chi_{--}\sim-\chi
_{++}\text{ \ \ \ \ , \ \ }\xi_{--}\sim\xi_{++}\text{ \ ,}\label{10}\\
\omega_{+-}  &  \sim0\text{ \ \ \ \ \ \ \ \ \ , \ \ }\chi_{+-}\sim0\text{
\ \ \ \ \ \ \ \ \ \ \ , \ \ }\xi_{+-}\sim0\text{ \ \ \ \ ,}\nonumber
\end{align}
Note that in the weak-amplification regime, the states $\left\vert
+\right\rangle $ and $\left\vert -\right\rangle $ couple to each other through
dynamical evolution, while in the strong-amplification regime each state
evolves independently. This fact represents an additional advantage of the
strong coupling regime, apart from the considerably stronger couplings that
require shorter atom-field interaction times for manipulations of the cavity
or the vibrational mode, making the dissipative mechanisms almost negligible.

\section{The Engineered Interactions}

Preparing the ion in the state $\left\vert i\right\rangle $, we obtain from
the Hamiltonian in Eq. (\ref{7}), returning to the Schr\"{o}dinger picture,
the result%
\begin{equation}
H=\omega a^{\dagger}a+\nu b^{\dagger}b+\left[  \omega_{ii}a^{\dagger}%
a+\chi_{ii}+\left(  \xi_{ii}\operatorname*{e}\nolimits^{-2i(\omega+\delta
)t}\left(  a^{\dagger}\right)  ^{2}+\mathrm{h.{c.}}\right)  \right]  \sin
^{2}\left[  \eta\left(  b^{\dagger}+b\right)  +\varphi\right]  \mathrm{{,}}
\label{11}%
\end{equation}
which will be analyzed in two cases corresponding to the ion centered at a
node or an anti-node of the standing wave, $\varphi=0$ or $\pi/2$,
respectively:
\begin{equation}
\sin^{2}\left[  \eta\left(  b^{\dagger}+b\right)  +\varphi\right]
\approx\left\{
\begin{tabular}
[c]{llllll}%
$1$, & adjusting & $\eta^{2}\ll1$ & and & $\varphi=\pi/2$ & ,\\
$\eta^{2}\left(  b^{\dagger}+b\right)  ^{2}$, & adjusting & $\eta^{4}\ll1$ &
and & $\varphi=0$ & .
\end{tabular}
\ \ \ \right.  \label{12}%
\end{equation}

Analyzing the case where $\sin^{2}\left[  \eta\left(  b^{\dagger}+b\right)
+\varphi\right]  \approx1$, in the interaction picture defined by the
transformation $U=\exp\left\{  -i\left[  \left(  \omega+\omega_{ii}\right)
a^{\dagger}a+\nu b^{\dagger}b\right]  t\right\}  $, we obtain, in both weak
and strong-amplification regimes, the first engineered Hamiltonian
\begin{equation}
\mathbf{H}_{1}=\xi_{ii}\left(  a^{\dagger}\right)  ^{2}+\mathrm{h.{c.}%
}\mathrm{{,}} \label{13}%
\end{equation}
where we have adjusted the cavity mode such that $\delta=\omega_{ii}$.
Interestingly enough, this Hamiltonian leads to the squeezing operator acting
only on the cavity mode: $S(\xi_{ii},t)=\exp\left[  -i\left(  \xi
_{ii}a^{\dagger2}+\xi_{ii}^{\ast}a^{2}\right)  t\right]  .$

Given that $\xi_{ii}=\left\vert \xi_{ii}\right\vert \operatorname*{e}%
\nolimits^{i\Theta}$, the degree of squeezing of the cavity field state
achieved through Hamiltonian $\mathbf{H}_{1}$ (Eq. (\ref{13})) is determined
by the factor $r(t)=2\left\vert \xi_{ii}\right\vert t$, while the squeeze
angle is given by $\Theta/2$. For a specific cavity mode and electronic
configuration of the trapped ion (i.e., for specific $\lambda_{1}$,
$\lambda_{2}$, and $\Delta$), the parameter $r(t)$ can be adjusted in
accordance with the coupling strength $\left\vert \Omega\right\vert $ and the
interaction time $t$. To estimate the degree of squeezing achieved we assume
trapped Rydberg atoms. Thus, considering typical cavity QED values for the
parameters involved, arising from Rydberg levels where the intermediate state
$\left\vert i\right\rangle $ is nearly halfway between $\left\vert
g\right\rangle $ and $\left\vert e\right\rangle $,\ with $\Delta\sim
3\times10^{6}$s$^{-1}$, we get $\left\vert \lambda_{1}\right\vert
\sim\left\vert \lambda_{2}\right\vert \sim3\times10^{5}$s$^{-1}$ \cite{BRH}.
In the weak-amplification regime, such values lead to $\delta=\omega
_{ii}=\omega_{w}\sim6\times10^{4}$s$^{-1}$, and assuming the coupling strength
$\Omega\sim3\times10^{5}$s$^{-1}$, we obtain $\left\vert \xi_{ii}\right\vert
\sim3\times10^{3}$s$^{-1}$. Therefore, for an ion-field interaction time about
$t\sim2\times10^{-4}$s, we get the squeezing factor $r(t)\sim1.2$ such that
the squeezing rate turns out to be $\mathcal{R}=\left(  1-\operatorname{e}%
^{-2r(t)}\right)  \times100\%\sim91\%$ (for an initial coherent state prepared
in the cavity). A laser pulse of longer duration leads to a squeezing rate
even greater than this remarkable rate (at the expense of intensifying the
dissipative mechanisms, neglected in the present work). Note that the
interaction time adopted here is one order of magnitude smaller than the decay
time of the open cavities used in cavity QED experiments \cite{BRH}.
Evidently, for the strong-amplification regime, we will obtain an even higher
degree of squeezing for the cavity mode.

We note that our cascade atomic-level scheme, where an auxiliary intermediate
state $\left\vert i\right\rangle $ is used to couple the dippole-forbidden
transition $\left\vert g\right\rangle \leftrightarrow\left\vert e\right\rangle
$, differs from the schemes used by both trapped ion groups: at NIST
\cite{NIST}, concentrated on Lambda configuration, and Innsbruck
\cite{Innsbruck}, where a dipole forbidden transition $\left\vert
g\right\rangle $ $\leftrightarrow$ $\left\vert e\right\rangle $ is induced by
applying a sufficiently strong electric field. However, the values presented
above for the atomic frequencies and couplings, arising from the Rydberg
levels used in Ref. \cite{BRH}, are around those considered in the Innsbruck configuration.

Next, to handle the case $\sin^{2}\left[  \eta\left(  b^{\dagger}+b\right)
+\varphi\right]  \approx\eta^{2}\left(  b^{\dagger}+b\right)  ^{2}$, it is
convenient to consider a picture defined by the transformation $U=\exp\left\{
-i\left[  \omega a^{\dagger}a+\Phi b^{\dagger}b\right]  t\right\}  $, where
the Hamiltonian reads
\begin{align}
\mathbf{H}  &  =\eta^{2}\left(  2b^{\dagger}b+1\right)  \left[  \omega
_{ii}a^{\dagger}a+\left(  \xi_{ii}\operatorname*{e}\nolimits^{-2i\delta
t}\left(  a^{\dagger}\right)  ^{2}+\mathrm{h.{c.}}\right)  \right]  +\eta
^{2}\left(  \omega_{ii}a^{\dagger}a+\chi_{ii}\right)  \left(
\operatorname*{e}\nolimits^{2i\Phi t}\left(  b^{\dagger}\right)
^{2}+\mathrm{h.{c.}}\right) \nonumber\\
&  +\eta^{2}\left(  \xi_{ii}\operatorname*{e}\nolimits^{-2i\left(  \delta
-\Phi\right)  t}\left(  a^{\dagger}\right)  ^{2}\left(  b^{\dagger}\right)
^{2}+\mathrm{h.{c.}}\right)  +\eta^{2}\left(  \xi_{ii}\operatorname*{e}%
\nolimits^{-2i\left(  \delta+\Phi\right)  t}\left(  a^{\dagger}\right)
^{2}\left(  b\right)  ^{2}+\mathrm{h.{c.}}\right)  , \label{14}%
\end{align}
and $\Phi=\nu+2\eta^{2}\chi_{ii}$. Evidently, varying the choice of the
detuning $\delta$ leads to distinct interactions, such that, by adjusting
$\delta$ to $\Phi$ and $-\Phi$, we obtain in both amplification regimes, after
rotating wave approximations, respectively
\begin{subequations}
\label{15}%
\begin{align}
\mathbf{H}_{2}  &  =\eta^{2}\omega_{ii}\left(  2b^{\dagger}b+1\right)
a^{\dagger}a+\eta^{2}\left(  \xi_{ii}\left(  a^{\dagger}\right)  ^{2}\left(
b^{\dagger}\right)  ^{2}+\mathrm{h.{c.}}\right)  \mathrm{{,}}\label{15a}\\
\mathbf{H}_{3}  &  =\eta^{2}\omega_{ii}\left(  2b^{\dagger}b+1\right)
a^{\dagger}a+\eta^{2}\left(  \xi_{ii}\left(  a^{\dagger}\right)  ^{2}\left(
b\right)  ^{2}+\mathrm{h.{c.}}\right)  \mathrm{{,}} \label{15b}%
\end{align}
where we have assumed, whatever the state of the cavity mode, the condition
$\Phi\gg\omega_{ii}\left\langle a^{\dagger}a\right\rangle +\chi_{ii}$,
$\xi_{ii}$. Under this same condition, but with $\left\vert \delta\right\vert
\ll\Phi$, also in both amplification regimes, we obtain the time-dependent
interaction%
\end{subequations}
\begin{equation}
\mathbf{H}_{4}=\eta^{2}\left(  2b^{\dagger}b+1\right)  \left[  \omega
_{ii}a^{\dagger}a+\left(  \xi_{ii}\operatorname*{e}\nolimits^{-2i\delta
t}\left(  a^{\dagger}\right)  ^{2}+\mathrm{h.{c.}}\right)  \right]
\mathrm{{.}} \label{16}%
\end{equation}

Finally, with $\left\vert \delta\right\vert \sim\left\vert \Phi\right\vert $
and $\left\vert \delta\pm\Phi\right\vert \sim\left\vert \Phi\right\vert $, or
$\delta\sim0$ and $\xi_{ii}\ll\omega_{ii}$,we obtain the Kerr-like
interaction
\begin{equation}
\mathbf{H}_{5}=\eta^{2}\omega_{ii}a^{\dagger}a\left(  2b^{\dagger}b+1\right)
\text{,} \label{16a}%
\end{equation}
which is suitable for introducing phases into one field state, according to
the intensity of the other.

\section{The Hamiltonian $\mathbf{H}_{4}$}

\subsection{Subcritical, Critical, and Supercritical Regimes}

To understand the behaviour of the system under the approximations leading to
Hamiltonian $\mathbf{H}_{5}$, its would be helpful to eliminate the
time-dependence of the interaction $\mathbf{H}_{4}$ through the unitary
transformation $U_{SC}(t)=\exp$ $\left[  -i\delta ta^{\dagger}a\right]  $. We
are left with the simplified form%
\begin{equation}
\mathbf{H}_{4}=\Xi(b^{\dagger}b)a^{\dagger}a+\frac{1}{2}\left(  \Gamma
(b^{\dagger}b)\left(  a^{\dagger}\right)  ^{2}+\mathrm{h.{c.}}\right)
\text{,} \label{17}%
\end{equation}
where the functions for the trapped-motion operators are given by
\begin{subequations}
\begin{align}
\Xi\left(  b^{\dagger}b\right)   &  =\eta^{2}\omega_{ii}\left(  2b^{\dagger
}b+1\right)  -\delta\text{,}\label{18a}\\
\Gamma(b^{\dagger}b)  &  =2\eta^{2}\xi_{ii}\left(  2b^{\dagger}b+1\right)
\text{.} \label{18b}%
\end{align}
In the Fock basis representation for the vibrational operators, the
$b^{\dagger}b$ operator is replaced by the motional excitation $m$ and the
Hamiltonian (\ref{17}) becomes%
\end{subequations}
\begin{equation}
\mathbf{H}_{4}(m)=\Xi(m)\left[  a^{\dagger}a+\frac{1}{2}\left(  \mathcal{F}%
(m)\left(  a^{\dagger}\right)  ^{2}+\mathrm{h.{c.}}\right)  \right]  \text{,}
\label{19}%
\end{equation}
where the function $\mathcal{F}(m)$ stands for the ratio
\begin{equation}
\mathcal{F}(m)=\frac{\Gamma(m)}{\Xi(m)}=\frac{2\xi_{ii}/\omega_{ii}}%
{1-\delta/\left[  2\omega_{ii}\left(  m+1/2\right)  \right]  } \label{19l}%
\end{equation}
Evidently, for $\Xi(m)=0$, we obtain the resonant amplification regime leading
to the maximum degree of squeezing of the cavity field. For $\Xi(m)\neq0$, the
absolute value $\left\vert \mathcal{F}(m)\right\vert $ determines three
different regimes of the non-resonant parametric amplification process, the
subcritical ($\left\vert \mathcal{F}\right\vert <1$), critical ($\left\vert
\mathcal{F}\right\vert =1$), and supercritical ($\left\vert \mathcal{F}%
\right\vert >1$) regimes. These regimes are characterized by oscillatory,
linear, and hyperbolic solutions of the Heisenberg equations of motion for the
evolution of the annihilation operator of the cavity mode given by%
\begin{equation}
a(t)=f(t)a-ig(t)a^{\dagger}\text{,} \label{20}%
\end{equation}
where the time-dependent functions in the subcritical , critical, and
supercritical regimes are
\begin{subequations}
\begin{align}
f_{<1}(t)  &  =\cos\left[  \mathfrak{w}(m)t\right]  -i\frac{\Xi(m)}%
{\mathfrak{w}(m)}\sin\left[  \mathfrak{w}(m)t\right]  \text{,}\label{21a}\\
f_{=1}(t)  &  =1-i\mathfrak{w}(m)t\text{,}\label{21b}\\
f_{>1}(t)  &  =\cosh\left[  \mathfrak{w}(m)t\right]  -i\frac{\Xi
(m)}{\mathfrak{w}(m)}\sinh\left[  \mathfrak{w}(m)t\right]  \text{,}
\label{21c}%
\end{align}
and%
\end{subequations}
\begin{subequations}
\begin{align}
g_{<1}(t)  &  =\frac{\Gamma(m)}{\mathfrak{w}(m)}\sin\left[  \mathfrak{w}%
(m)t\right]  \text{,}\label{22a}\\
g_{=1}(t)  &  =\Gamma(m)t\text{,}\label{22b}\\
g_{>1}(t)  &  =\frac{\Gamma(m)}{\mathfrak{w}(m)}\sinh\left[  \mathfrak{w}%
(m)t\right]  \text{,} \label{22c}%
\end{align}
where $\mathfrak{w}^{2}(m)=\left\vert \left\vert \Gamma(m)\right\vert ^{2}%
-\Xi^{2}(m)\right\vert $. Evidently, for $\left\vert \mathcal{F}\right\vert
>>1$, we are close to the resonant regime. Each of these regimes results in a
different squeezing process of the cavity-field state, as already discussed in
Refs. \cite{Celso,Salomon}. However, since in the present model the
vibrational field is an additional ingredient, for fixed values of $\delta$,
$\eta$, $\omega_{ii}$, and $\xi_{ii}$, the various regimes can be achieved by
manipulating the excitationnumber $m$ of the vibrational mode, except in some
particular cases where the adjustment of the detuning $\delta$ results in a
fixed amplification regime for all values of $m$. These are the cases of Fig.
2 (a), where $\delta=0$ leads to the constant function $\mathcal{F}=2\xi
_{ii}/\omega_{ii}$, and Fig. 2(b), where $\delta$ is adjusted in such a way
that $\left\vert \mathcal{F}\right\vert <1$ or $\left\vert \mathcal{F}%
\right\vert >1$ for all $m$. In Fig. 2 (c), the parameters are adjusted to get
an inversion of the behavior of function $\mathcal{F}$, from $\left\vert
\mathcal{F}\right\vert \lessgtr1$ to $\left\vert \mathcal{F}\right\vert
\gtrless1$, passing or not through $\left\vert \mathcal{F}\right\vert =1$. We
note that in Figs. 2 (b) and (c) the decreasing functions are singular for
$m=0$ where we have the resonant amplification regime. In Figs. 2 (a), (b) and
(c) $\left\vert \mathcal{F}\right\vert $ behaves nonmonotonically. With the
parameters of Fig. 2 (d), we obtain two different behaviors with the same
$\omega_{ii}$: for $\delta=20\omega_{ii}$, we start from the subcritical
regime (passing or not through the critical regime with a suitable adjustment
of $\delta$), while for $\delta=2\omega_{ii}$ we begin from the critical
regime at $m=0$. In both cases, the critical regime is reached assymptotically
from the supercritical regime. In Fig. 2 (e) we have the subcritical regime
for all values of $m$ except for $m=10$, at which the supercritical regime is
found. Finally, in Fig. 2(f) we have the same behavior as in Fig. 2 (e),
except that for $m=10$ we have a singularity, indicating the resonant regime
for this value of $m$.

In the case where the vibrational field is prepared in a coherent state
$\beta$, it is possible to choose the mean excitation $\left\vert
\beta\right\vert ^{2}$ in such a way that all the significant values for its
Fock components $m$ lie in the subcritical or supercritical region. As an
example, within the parameters of Fig. 2 (c) and a coherent state $\beta\sim
4$, we obtain the supercritical (subcritical) regime for all the significant
values of $m$ when $\omega_{ii}=\left(  10/9\right)  \left\vert \xi
_{ii}\right\vert $ and $\delta=-2\omega_{ii}$ ($\omega_{ii}=\left(
10/4\right)  \left\vert \xi_{ii}\right\vert $ and $\delta=0.5\omega_{ii}$).

\subsection{A Fock State Filter}

Let us consider $\delta=\omega_{ii}(2M+1)$, where the resonant regime occurs
only for $m=M$ and the subcritical regime at all other values of $m$, as in
Fig. 2 (f). Starting from the cavity mode in the vacuum state and the
vibrational mode in a coherent state $\left\vert \beta\right\rangle
=\sum\nolimits_{m}C_{m}\left\vert m\right\rangle $, we obtain from Hamiltonian
$\mathbf{H}_{4}(m)$ the evolved superposition%
\end{subequations}
\begin{equation}
\left\vert \psi(t)\right\rangle =C_{M}\left\vert M\right\rangle S_{M}%
(t)\left\vert 0\right\rangle +\sum\limits_{\substack{m=0\\(m\neq M)}}^{\infty
}C_{m}\left\vert m\right\rangle S_{m}(t)\left\vert 0\right\rangle \text{,}
\label{23}%
\end{equation}
where $S_{M}(t)$ $=\exp\left[  -i\left(  \Gamma(M)\left(  a^{\dagger}\right)
^{2}+\mathrm{h.{c.}}\right)  t/2\right]  $ stands for the ideal squeezing
operator and $S_{m}(t)=\exp\left\{  -i\mathbf{H}_{4}(m)t\right\}  $ indicates
the nonresonant squeezing operator in the subcritical regime. Adjusting
$\omega_{ii}\gg\left\vert \xi_{ii}\right\vert $, such that $\mathfrak{w}%
(m)\sim\left\vert \Xi(m)\right\vert $ and $\left\vert \Gamma(m)/\Xi
(m)\right\vert \sim\left\vert \xi_{ii}\right\vert /\omega_{ii}\ll1$, the
squeezing process is strongly nonresonant for all values of $m$ other than
$M$. Consequently, there will be practically no photon injection into the
cavity mode from the nonresonant squeezing ($NS$) process, since
\begin{align}
\left\langle a^{\dagger}a\right\rangle _{NS}  &  =\left(  1-\left\vert
C_{M}\right\vert ^{2}\right)  ^{-1}\sum\limits_{\substack{m=0\\(m\neq
M)}}^{\infty}\left\vert C_{m}\right\vert ^{2}\left\langle 0\right\vert
S_{m}^{\dagger}(t)a^{\dagger}aS_{m}(t)\left\vert 0\right\rangle \nonumber\\
&  =\left(  1-\left\vert C_{M}\right\vert ^{2}\right)  ^{-1}\sum
\limits_{\substack{m=0\\(m\neq M)}}^{\infty}\left\vert C_{m}\right\vert
^{2}\left\vert \frac{\Gamma(m)}{\mathfrak{w}(m)}\right\vert ^{2}\sin
^{2}\left[  \left\vert \mathfrak{w}(m)\right\vert t\right]  \lesssim
\frac{\left\vert \xi_{ii}\right\vert }{\omega_{ii}} \label{24}%
\end{align}
Therefore, for such a nonresonant process, the cavity field remains close to
the vacuum state. On the other hand, the resonant squeezing ($RS$) process
accounts for a significant photon injection into the cavity mode, whose
excitation becomes%
\begin{equation}
\left\langle a^{\dagger}a\right\rangle _{RS}=\left\langle 0\right\vert
S_{M}^{\dagger}(t)a^{\dagger}aS_{M}(t)\left\vert 0\right\rangle =\sinh
^{2}\left[  \left\vert \Gamma(M)\right\vert ^{2}t\right]  \label{25}%
\end{equation}
Therefore, after a convenient time interval $t$, when $\left\langle
a^{\dagger}a\right\rangle _{RS}$ is appreciably larger than unity, a
measurement of the cavity field, in a state with a considerable number of
photons, will project the vibration mode into the Fock state $\left\vert
M\right\rangle $. Evidently, the probability of success in generating the
number state $\left\vert M\right\rangle $, given by $\left\vert C_{M}%
\right\vert ^{2}=\operatorname{e}^{-\left\vert \beta\right\vert ^{2}%
}\left\vert \beta\right\vert ^{2M}/M!$, can be maximized by adjusting the
amplitude of the coherent state $\beta$. This measurement is accomplished by
passing through the cavity a stream of ground-state two-level atoms
interacting resonantly with the cavity mode \cite{Brune}.

\subsection{The Semiclassical Approximation}

Next, we analyse Hamiltonian (\ref{16}) for the supercritical case and the
semiclassical approximation where the annihilation (creation) operator $b$
($b^{\dagger}$) is replaced by the amplitude $\beta$ ($\beta^{\ast}$) of a
strong coherent state. (Note that this procedure must be carried out from the
initial model in Eq. (\ref{2}), so that the factor $\left(  2b^{\dagger
}b+1\right)  $ in Eq. (\ref{16}) should be replaced by $\left\vert
\beta\right\vert ^{2}$.) Adjusting the detuning between the cavity field and
the atom such that $\delta=2\eta^{2}\left\vert \beta\right\vert ^{2}%
\omega_{ii}$, the unitary transformation $U_{SC}(t)=\exp$ $\left[  -i\delta
ta^{\dagger}a\right]  $ allows us to rewrite Hamiltonian $\mathbf{H}_{5}$ in
its semiclassical form%
\[
\mathbf{H}_{SC}=2\eta^{2}\left\vert \beta\right\vert ^{2}\left(  \xi
_{ii}\left(  a^{\dagger}\right)  ^{2}+\mathrm{h.{c.}}\right)  \text{,}%
\]
which describes an ideal squeezing process with the squeezing factor given by
$r=4\eta^{2}\left\vert \beta\right\vert ^{2}\left\vert \xi_{ii}\right\vert t$.
Therefore, the larger the amplitude $\beta$, the shorter the time required to
attain a given degree of squeezing. To estimate the validity of this
approximation we compare the variance of the squeezed quadrature $\left(
\Delta X_{sq}\right)  ^{2}$ computed from the two Hamiltonians, Eqs.
(\ref{16}) -- under the same transformation $U_{SC}$ -- and (\ref{17}). In
Fig. 3 we plot $\left(  \Delta X_{sq}\right)  ^{2}$ against the squeezing
factor $r$, where $X_{sq}\equiv(a+a^{\dagger})/2$, for different values of the
coherent vibrational state $\beta$, assumed to be real. Since the abscissa
represents the squeezing factor $r$, the straight line in Fig. 3 describes the
evolution of variance $\left(  \Delta X_{sq}\right)  ^{2}$ governed by the
semiclassical Hamiltonian $\mathbf{H}_{SC}$, for any value $\beta>0$. On the
other hand, the dotted, dashed-dotted, and dashed lines describe $\left(
\Delta X_{sq}\right)  ^{2}$ for the full quantum Hamiltonian $\mathbf{H}_{5}$,
taking $\beta=1$, $5$, and $10$ , respectively. As expected, the larger the
value of the amplitude of the vibrational field $\beta$,\ the better the
semiclassical approximation. For $\beta=10$, the variance obtained through the
semiclassical interaction fits the quantum description to a good approximation
for a degree of squeezing about $90\%$, i.e., $r\sim1$, which is achieved in a
time interval about $4\times10^{2}\times\eta^{2}\left\vert \xi_{ii}\right\vert
^{-1}$s.

\section{Conclusions}

In this paper we have presented a scheme for engineering quadratic and
bi-quadratic phonon-photon interactions through a driven three-level ion
inside a cavity. The adiabatic approximation was employed and new aspects of
this technique were revealed through our analysis. Umtil now, this
approximation has been applied only in the weak-amplification regime, where
the atom-field coupling parameters are considerably smaller than their
detunings. In our approach, we also considered the strong-amplification regime
where, in contrast, the detunings were made considerably smaller than the
atom-field coupling parameters.

A detailed analysis of the squeezing process of the cavity mode was
accomplished by considering a particular phonon-photon interaction, described
by Hamiltonian (\ref{16}), revealing the possibility of one resonant and three
different nonresonant regimes of parametric amplification of the cavity mode.
Interestingly enough, such regimes of parametric amplification can be
modulated through the excitation of the vibrational-field state. We also
showed how to generate, through the same Hamiltonian (\ref{16}), a filter of
any Fock state for the vibrational mode via a projective measurement of the
cavity-field state.

Finally, we presented a detailed analysis of the semiclassical approximation
which enabled us to replace the operators describing the vibrational field in
Hamiltonian (\ref{16}) by classical amplitudes, simplifying considerably this
interaction. The same analysis of the validity of the semiclassical regime can
be carried out for Hamiltonians (\ref{15}). It is worth mentioning a recent
achievement by G. R. Guth\"{o}hrlein et al. \cite{Nature}, where a near-field
probe with atomic-scale resolution, a single calcium ion in a radio-frequency
trap, is reported. This work opens the way for performing higher resolution
cavity quantum electrodynamics experiments with a single trapped particle.

\textbf{Acknowledgments}

We wish to express thanks for the support from PIADRD/UFSCar, FAPESP, CAPES,
and CNPq (Instituto do Mil\^{e}nio de Informa\c{c}\~{a}o Qu\^{a}ntica),
Brazilian agencies. We also thank Dr. R. M. Serra for helpful discussions.

\textbf{Figures Caption}

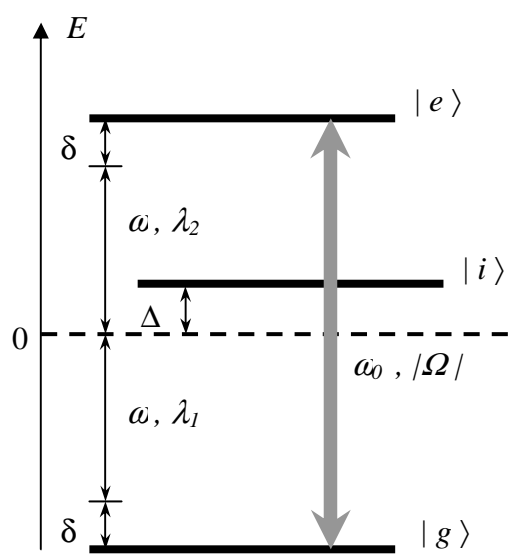
Fig. 1. Energy diagram of the three-level trapped ion in the ladder configuration.

Fig. 2. Distinct forms of behavior of $\mathcal{F}(m)$, for different values
of $\omega_{ii}\ $and $\delta$, which determines the regimes of the parametric
amplification processes.

Fig. 3. The variance of the squeezed quadrature $\left(  \Delta X_{sq}\right)
^{2}$ against the squeezing factor $r$ for the semiclassical Hamiltonian
$\mathbf{H}_{SC}$ and for the full quantum Hamiltonian $\mathbf{H}_{5}$, for
$\beta=1$, $5$, and $10$.


\begin{thebibliography}{99}
\bibitem {Forbidden}This dipole-forbidden transition can be induced by taking
advantage of dipole-allowed transitions to another auxiliary level . In this
case, a fourth level $\left\vert f\right\rangle $ must be added to the atomic
system together with appropriate classical fields inducing a Raman transition
with effective couplings given by $\Omega=g_{1}^{\ast}g_{2}/\widetilde{\Delta
}$, $g_{1}$ and $g_{2}$ being the coupling constants in the vicinity of the
dipole-allowed transitions $\left\vert g\right\rangle $ $\leftrightarrow$
$\left\vert f\right\rangle $ and $\left\vert e\right\rangle $ $\leftrightarrow
$ $\left\vert f\right\rangle $, both with detuning $\widetilde{\Delta}\gg$
$\left\vert g_{1}\right\vert $,$\left\vert g_{2}\right\vert $. For the
strong-coupling regime, it is required that $\Omega\sim10^{6}$s$^{-1}$, which
follows from $g_{1}\sim g_{2}\sim10^{7}$s$^{-1}$ (easily achieved for
dipole-allowed transitions). Note that we may also consider the auxiliary
level $\left\vert i\right\rangle $ itself for this purpose, taking care to
keep the classical fields far from resonance with the cavity modes.
\end{thebibliography}
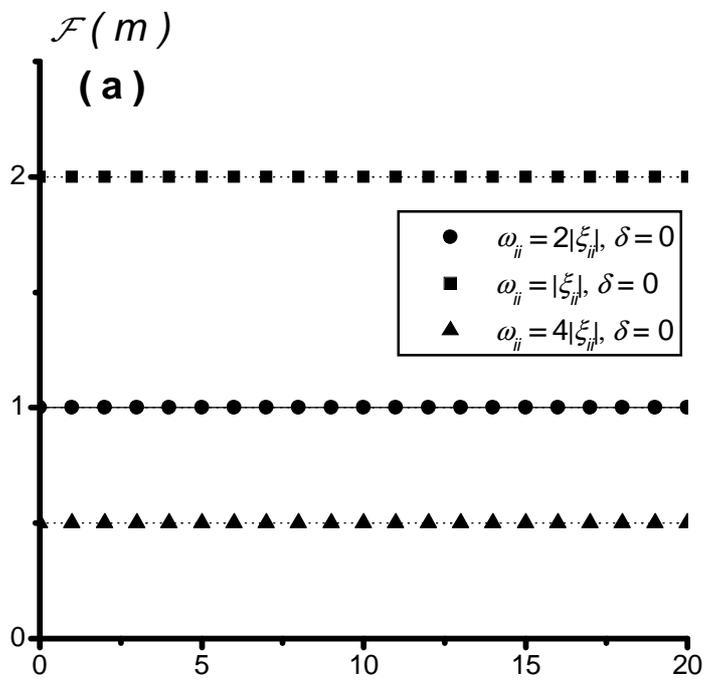
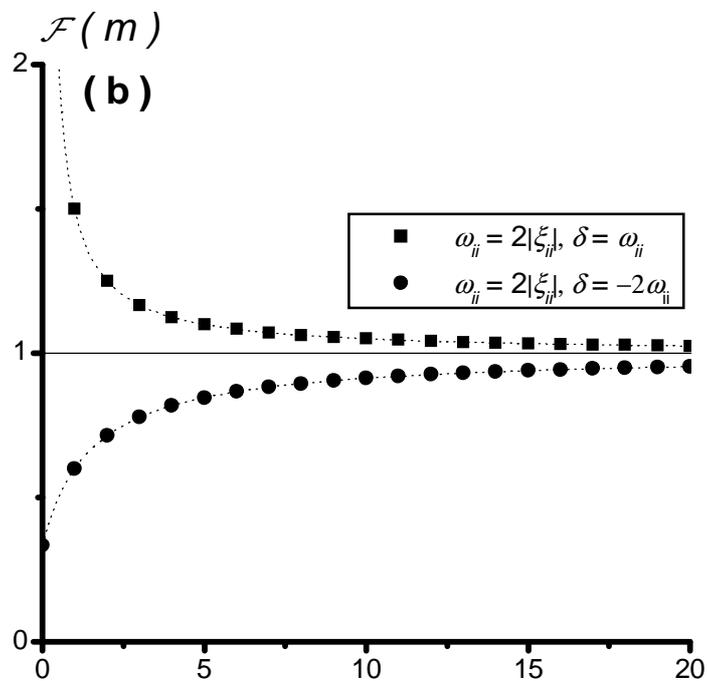
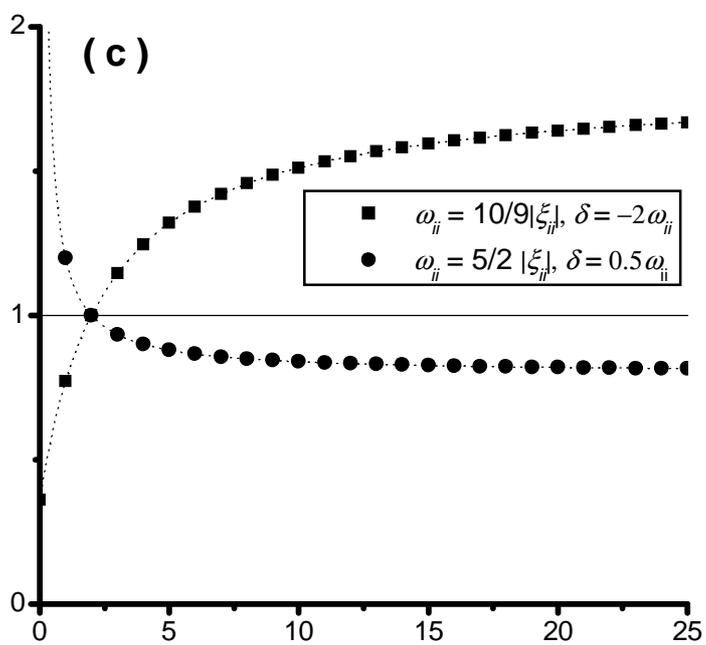
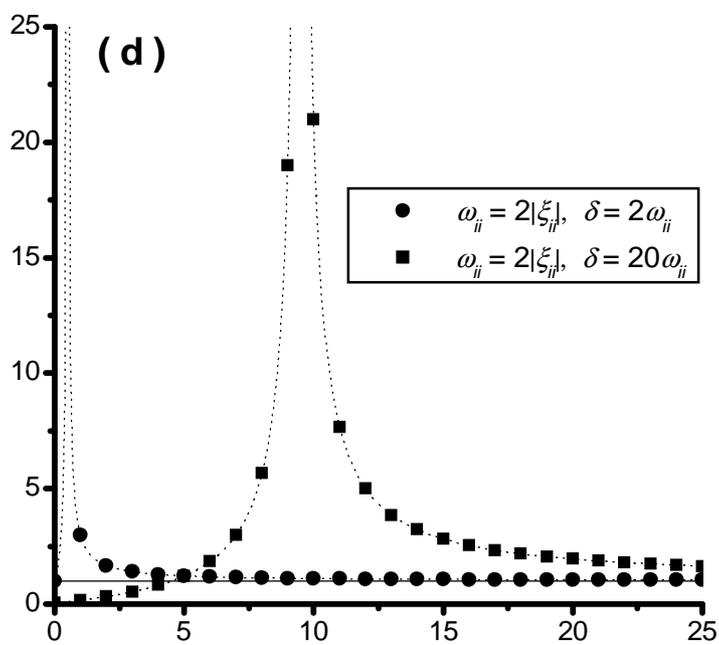
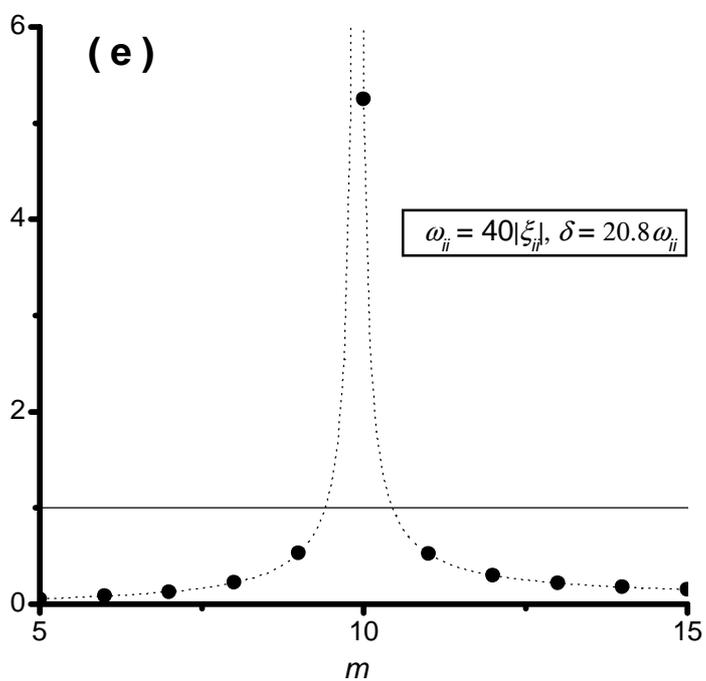
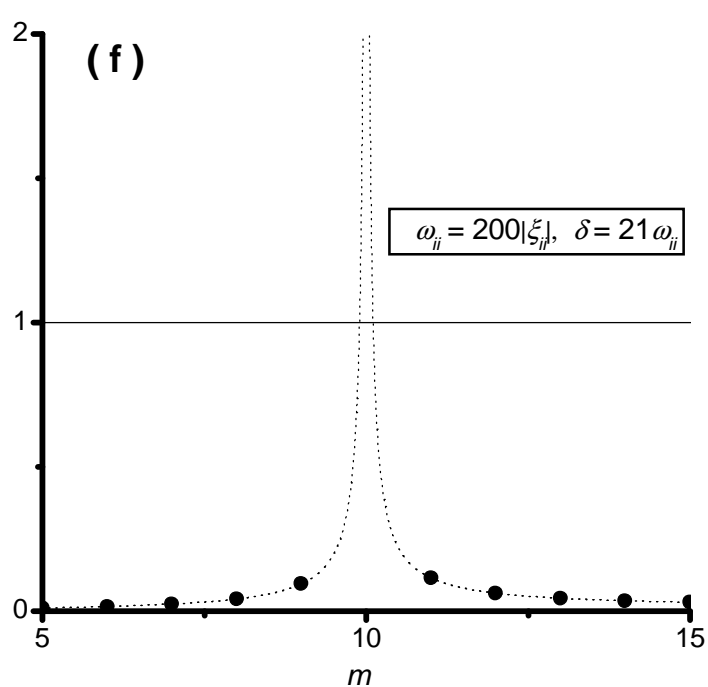
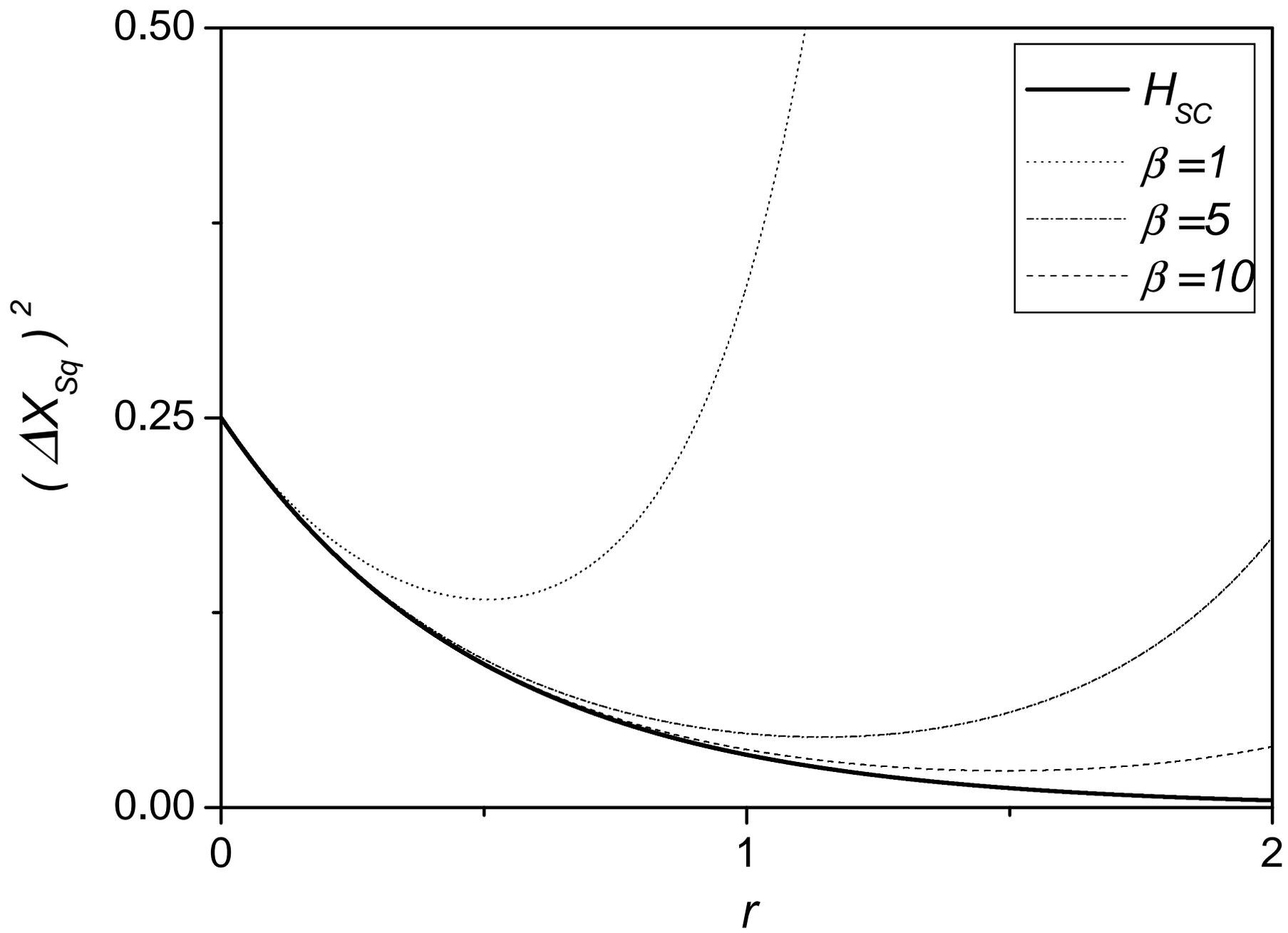
\end{document}